# Magnon Diffusion Length and Longitudinal Spin Seebeck Effect in Vanadium Tetracyanoethylene (V[TCNE]$_x$, $x \sim 2$)

Seth W. Kurfman[1*], Denis R. Candido[2*], Brandi Wooten[3], Yuanhua Zheng[4], Michael J. Newburger[1], Shuyu Cheng[1], Roland K. Kawakami[1], Joseph P. Heremans[1,3,4], Michael E. Flatté[2,5], and Ezekiel Johnston-Halperin[1]

[1]Department of Physics, The Ohio State University, Columbus, OH 43220, USA

[2]Department of Physics and Astronomy, The University of Iowa, Iowa City, IA 52242, USA

[3]Department of Materials Science and Engineering, The Ohio State University, Columbus, OH 43220, USA

[4]Department of Mechanical Engineering, The Ohio State University, Columbus, OH 43220, USA

[5]Department of Applied Physics, Eindhoven University of Technology, Eindhoven, The Netherlands

[*]These authors contributed equally

**Spintronic, spin caloritronic, and magnonic phenomena arise from complex interactions between charge, spin, and structural degrees of freedom that are challenging to model and even more difficult to predict. This situation is compounded by the relative scarcity of magnetically-ordered materials with relevant functionality, leaving the field strongly constrained to work with a handful of well-studied systems that do not encompass the full phase space of phenomenology predicted by fundamental theory. Here we present an important advance in this coupled theory-experiment challenge, wherein we extend existing theories of the spin Seebeck effect (SSE) to explicitly include the temperature-dependence of magnon non-conserving processes. This expanded theory quantitatively describes the low-temperature behavior of SSE signals previously measured in the mainstay material yttrium iron garnet (YIG) and predicts a new regime for magnonic and spintronic materials that**

have low saturation magnetization, $M_S$, and ultra-low damping. Finally, we validate this prediction by directly observing the spin Seebeck resistance (SSR) in the molecule-based ferrimagnetic semiconductor vanadium tetracyanoethylene (V[TCNE]$_x$, $x$ ~2). These results validate the expanded theory, yielding SSR signals comparable in magnitude to YIG and extracted magnon diffusion length ($\lambda_m > 1\ \mu m$) and magnon lifetime for V[TCNE]$_x$ ($\tau_{th} \approx 1 - 10\ \mu s$) exceeding YIG ($\tau_{th} \sim 10\ ns$). Surprisingly, these properties persist to room temperature despite relatively low spin wave stiffness (exchange). This identification of a new regime for highly efficient SSE-active materials opens the door to a new class of magnetic materials for spintronic and magnonic applications.

## INTRODUCTION

The interconnection and conversion between heat, charge, and spin currents is central to the study of spin caloritronics, a subfield of spintronics[1]–[5]. As thermal processes and temperature affect numerous material parameters, including magnetization and (quasi-)particle transport, spin-thermal physics can be a powerful probe of fundamental mechanisms in magnetic materials. However, this complexity also makes it particularly difficult to parse the numerous interactions that drive these spin-thermal interactions. For example, the spin Seebeck effect (SSE) has drawn significant attention from the spin caloritronic community, and has led to competing and conflicting views of the mechanisms responsible for the effect[6]–[13]. Since its initial discovery[14], the SSE has been measured in a variety of materials, including ferro(ferri)magnets[15]–[17], paramagnets[18]–[21], antiferromagnets[22]–[24], graphene/graphyne nanoribbons[25], a quantum magnet candidate[26], and even a non-magnetic material[27], and can be observed in metallic, semiconducting, and insulating magnetic materials[4]. Recently, a nuclear contribution of the SSE has also been reported[28]. While these material systems all produce measurable SSE signals, their electronic and magnetic properties differ and thus complicate a comprehensive understanding of the contributing effects. This variability highlights the need for a more complete theory of the underlying mechanism and experimental signatures observed in these spin-thermal experiments.

Here, we describe the mechanisms of the SSE based on a bulk magnon spin current for ferro-(ferri-)magnetic materials[7], [13], [24], [29], emphasizing the magnon interactions and their dependencies that are required for a theoretical model to accurately predict and explain SSE physics. Our theory quantitatively predicts the temperature-dependence of SSE signals, including the temperature for maximum SSE efficiency as a function of the sample thickness, measured in previous YIG SSE studies[30], [31]. We show this provides new insights into ideal magnetic material characteristics and parameters for magnonic spin injection devices, identifying optimized regimes of saturation magnetization, magnon lifetime and scattering time, thermal conductance, and exchange stiffness. Specifically, this theory predicts a regime for SSE-active materials that have low saturation magnetization, $M_S$, and ultra-low damping. We validate this prediction by performing systematic SSE measurements in the low-loss and low-$M_S$ ferrimagnet vanadium tetracyanoethylene (V[TCNE]$_x$, $x$ ~2). We find our theoretical model accurately predicts the large SSE response observed in these V[TCNE]$_x$ thin films and allows for the extraction of the magnon diffusion coefficient ($D_m > 0.002\ cm^2/s$), magnon lifetime ($\tau_{th} \approx 1 - 10\ \mu s$), and magnon diffusion length ($\lambda_m > 1\ \mu m$). Surprisingly, these properties persist to room temperature despite the low value of the spin wave stiffness (exchange) extracted for these materials. These results provide new criteria and guidance for the search for spintronic and magnonic materials and validate the class of magnetic coordination compounds to which V[TCNE]$_x$ belongs as a fruitful direction for further development and exploration[32].

**RESULTS**

The spin Seebeck effect is currently understood as the combination of two distinct processes: the generation of a non-equilibrium spin current at a ferromagnet/non-magnetic metal (FM/NM) interface[33], [34], and the conversion of that spin current into a transverse charge current in the NM via the inverse spin Hall effect (ISHE)[35] (see Fig. 1(a)). Early models suggested the spin current injection arises from a difference in temperature between magnons and electrons at the interface[6], [8]–[10], though later descriptions predicted the spin current is caused by a temperature gradient in the magnon excitations in the bulk of the FM[7], [13]. Recent time-resolved studies of these processes indicate that in fact the latter process yields the dominant contribution to the temperature-dependence of the injected spin current whereas the former contribution is relatively constant with temperature[36]–[39] – hence, in this work the bulk magnon spin current model is used to include temperature-dependent processes in the SSE. Critically, prior work considering the bulk magnon spin current has modeled the resulting spin flow using an explicit temperature-dependence for magnon-conserving scattering processes but considered an effective, constant scattering rate for *non-conserving* magnon processes[7], [24], [29]. Although these earlier models effectively capture the general trends of the spin current with temperature and FM thickness, the temperature dependences of non-conserving processes cannot be neglected as they are critical to a theory that more fully and quantitatively describes the quasiparticle dynamics implicated in these dependencies. Explicitly including the temperature-dependence of non-conserving processes leads to the following expression for the spin current incident at the FM/NM interface due to an acoustic magnon branch[7], [29] (see Supplemental Information for details),

$$j_m^y(z=d) = -F \frac{B_1 B_S}{\sqrt{B_0 B_2}} \tanh\left(\frac{d}{2\lambda_m}\right) g_r^{\uparrow\downarrow} \frac{\partial T}{\partial z} \quad (1)$$

where $B_0$, $B_1$, $B_2$ and $B_S$ are integrals involving the Bose-Einstein distribution, group velocity of magnons, and magnon lifetimes; $\lambda_m$ is the magnon diffusion length, $g_r^{\uparrow\downarrow}$ is the the real part of the spin-pumping conductance, and $\partial T/\partial z$ is the temperature gradient along the magnetic material. The coefficient $F$ is given by the expression $F = \gamma \hbar k_B k_{ZB}^2 \omega_{ZB} \sqrt{\tau_{th}\tau_0}/2\pi\sqrt{3}4\pi M_S$, with $4\pi M_S$ the saturation magnetization (in cgs units), $\gamma$ the gyromagnetic ratio, $k_{ZB}$ the wavevector defining the first Brillouin zone for the acoustic magnon branch with corresponding energy $\omega_{ZB}$ [See Fig. 1(b)], $\tau_0 = \eta_q \tau_m$ is the relaxation time of $\boldsymbol{k} = 0$ magnons with $\eta_q$ the wavevector dependence of the relaxation time, and $\tau_m = \tau_m(T)$ ($\tau_{th} = \tau_{th}(T)$) is the temperature-dependent relaxation time of processes conserving (not conserving – also known as the magnon lifetime) the total number of magnons (see Supplement for details on the temperature-dependences of $\tau_m$ and $\tau_{th}$). Once the spin current enters the NM (in this case, Pt), it is converted into an electric current via the inverse spin Hall effect[35] (see Fig. 1(a)). This current produces a voltage difference across the Pt metal given by[7], [29],

$$V_{SSE} = R_{Pt} l_w \lambda_N \theta_{SH} \tanh\left(\frac{t_{Pt}}{2\lambda_N}\right) j_m^y(z=d) \quad (2)$$

where $R_{Pt}$ is the resistance of the Pt metal, $l_w$ ($t_{Pt}$) is its width (thickness), $\lambda_N$ the electronic spin diffusion length of Pt, $\theta_{SH}$ the spin Hall angle, and $e$ the electronic charge. It is critical to highlight that this voltage is inversely proportional to the saturation magnetization $4\pi M_S$, thus indicating that materials with low-$M_s$ values may demonstrate highly efficient spin injection and should be explored more thoroughly for such applications.

As this theory assumes that the spin current is created by the non-equilibrium acoustic bulk magnons, for a given thermal gradient the spin current penetrating into the Pt layer is strongly dependent on the (magnon) temperature. That is, three general regimes provide different temperature-dependent magnon excitation behaviors. For low temperatures such that the thermal energy is much smaller than the magnon gap, $k_B T \ll \gamma H$, zero spin current is expected due to the lack of thermally excited magnons which have energies above the energy gap. With increasing temperature, one expects a rapid increase of the spin current as the thermal energy becomes sufficient to excite the relevant magnon band, that is $\gamma H \lesssim k_B T \lesssim \hbar \omega_{ZB}$. The spin current then increases slowly for intermediate temperature $T \gtrsim T_{eff}$ with $T_{eff} = \hbar \omega_{ZB}/k_B$, until it saturates for temperature $k_B T \gg \hbar \omega_{ZB}$, as the net number of magnons carrying the spin current also saturates (i.e. the *net* magnon number $\delta n$ above the equilibrium population saturates at higher temperature – see Supplement). This trend is represented by the magenta dashed line in Fig. 1(c). Accordingly, the temperature values for which the spin current saturates are strictly dependent on $\omega_{ZB}$, characterizing the magnon energy bandwidth (see Figs. 1(b) and (c)).

Despite the accuracy of the discussion above regarding the thermal excitation of magnons, it only describes real behavior if the magnon parameters $\tau_{th}, \tau_m$, and $M_S$ contained in Eq. (1) remained constant as a function of the temperature for the experimentally relevant temperature range. However, it is well known that both $\tau_m$ and $M_S$ decrease with increasing temperature[40], [41]. Accordingly, in the previous theory the inclusion of the temperature dependence on $\tau_m$ leads to a peak of the spin current at $T = T_{peak}$, which strongly depends on the scattering time as a function of temperature and on $T_{eff}$. With increasing temperatures above $T_{eff}$ the magnon lifetime dramatically reduces, consequently reducing the SSE signal as summarized by Fig. 1(d). Further, the literature also reports a temperature dependence in the magnon relaxation time, $\tau_{th}$, [40], [42] that has not previously been included in theoretical models of SSE[7], [29]. In this work, this temperature dependence is included explicitly, and the results directly and quantitatively compared to experimental results.

This theoretical model further accounts for the general dependence/role of the FM thickness as shown in Fig. 1(e), where different thicknesses produce the same profile, differing only by the amplitude. This can be understood by observing that the $\tanh(d/2\lambda_m)$ term in Eq. 1 is approximately proportional to $d$ for $d \ll 2\lambda_m$. However, it is also important to address the fact that both magnon lifetime and magnon scattering time also depend on the thickness of the film[40], [43]. This was first explained in Refs. [40, 43] where imperfections (pits) along the surface of the magnetic material act as scattering centers for magnons. Accordingly, a proper assessment of the SSE in magnetic materials also needs to consider the thickness dependence within the magnetic material properties. Here, the theory developed in Refs. [7, 39] has been extended by including the thickness dependence of magnon scattering times for $\tau_m$ which, when using parameters for YIG, accurately describes and predicts the thickness dependent features seen in prior YIG SSE studies[30], [44], [45] with better agreement than previous theory calculations[24] (see Supplement).

The excellent agreement between the model presented above and experimental data provides strong evidence for extending this theory to other magnetic materials, thus providing a pathway to accurately predict optimal material parameters for efficient (thermally-excited) spin pumping processes. Specifically, as mentioned above the theory predicts efficient spin pumping for magnetic materials with low-$M_S$. Additionally, magnetic materials with low Gilbert damping, which can produce magnons with longer lifetimes, further enhance spin pumping effects. The above theory also appears to suggest large magnon bandwidth and large exchange stiffness are ideal for such applications; however, this prediction is inaccurate as the spin lifetimes and scattering rates increase with increasing magnon bandwidth and exchange stiffness. Therefore, this highlights the competition between the parameters that govern the magnitude of the injected spin current. To date, such spin injection studies have focused on magnetic materials with large exchange stiffness, wide magnon bandwidth, and high saturation magnetization. It should be noted that since the magnetic ordering temperature is proportional to the spin wave stiffness a small exchange stiffness implies a low magnetic ordering temperature, and hence efficient spin injection only at low temperatures. These results suggest the opportunity to explore a new phase space of magnetic materials with small exchange stiffness, narrow magnon bandwidth, and low saturation magnetization for applications in SSE-based spintronics and magnonics.

In this context, the low-loss ($\alpha \sim 4 \times 10^{-5}$) and low-$M_S$ ($4\pi M_{eff} \sim 100$ G) organic-molecule-based semiconducting ferrimagnetic material V[TCNE]$_x$ provides an appealing testbed for evaluating this prediction. This ferrimagnetic material, with damping comparable to high-quality YIG films, has attracted significant interest from the spintronics, magnonics, and quantum information science and engineering (QISE) communities due to its low-loss and highly coherent magnetic resonance properties. V[TCNE]$_x$ ferromagnetic resonance (FMR) shows remarkably narrow linewidths ($< 1\ Oe$) at X-band frequencies (9.8 GHz) with $Q$-factors exceeding 3,000 in thin films[42], [46], [47] and spin wave $Q$-factors over 8,000 in patterned V[TCNE]$_x$ microstructures[48], while thin films maintain low-loss resonance at cryogenic temperatures[42]. Further, spin waves in V[TCNE]$_x$ thin films have been measured propagating over cm distances[49]. These low-loss magnonic properties are particularly surprising as V[TCNE]$_x$ films lack long-range structural order[47], which typically increases magnon scattering and damping in magnetic materials.

To test the validity of the theory presented above, V[TCNE]$_x$ SSE devices are fabricated in the longitudinal SSE (LSSE) configuration (Fig. 1(a)). An air-free transfer method is used between Pt deposition on Al$_2$O$_3$ substrates via molecular beam epitaxy (MBE) and V[TCNE]$_x$ deposition via chemical vapor deposition (CVD) in order to produce a high-quality interface between the Pt and V[TCNE]$_x$ layers, resulting in highly reproducible SSE devices (see Supplement). A typical device response is shown in Figure 2(a), where the ISHE voltage in the Pt layer is measured as a function of the external magnetic field for multiple heater powers (i.e. different thermal gradients across the V[TCNE]$_x$ layer). As discussed above, this ISHE voltage is a direct measurement of the injected spin current across the V[TCNE]$_x$/Pt interface, which is in turn dependent on the temperature gradient generated by the heat flow from the heater. The voltage measured in the Pt layer is linear with the applied heater power, as expected (see Supplement). When the magnetic field is varied, the ISHE voltage tracks the magnetization of the V[TCNE]$_x$, where the sign change of the voltage is a result of the cross product $\widehat{E}_{ISHE} \propto \hat{j}_m \times \widehat{M}$ where $\widehat{M}$ is the unit vector describing the magnetization direction[50]. However, reporting only the SSE voltage is not a sufficient benchmark for the strength of the SSE in a given material, as the voltage scales extrinsically with the device size and with the applied thermal gradient $\partial T/\partial z$ as described in Eq (2). Accordingly, the spin Seebeck resistance (SSR) is a more appropriate and consistent metric[51], [52] for these measurements which normalizes the electrical signal to the input heat flux[45]

$$SSR = \frac{E_{Pt}}{j_Q} = \frac{V_{ISHE} * w}{P}, \qquad (3)$$

where $E_{Pt} = \frac{V_{ISHE}}{l_w}$ is the electric field in the Pt detection layer and $j_Q = \frac{P}{A}$ is the thermal flux from the heater passing through the V[TCNE]$_x$ area $A = l_w \times w$.

The SSR signal is shown in Figure 2(b) and (c) for 100 nm, 400 nm, 800 nm, and 1200 nm V[TCNE]$_x$ films, where the SSR is plotted with different *x*-axes (temperature and thickness, respectively) to clarify specific features. The SSR for V[TCNE]$_x$ films ranges from 10-60 nm/A, consistent with values seen in YIG films of comparable thickness[45], [52], [53] (also included in Figure 2(b)). The SSR increases with decreasing temperature seen in Figure 2(b), which is consistent with increasing $\lambda_m$ and enhanced $\tau_m$ at lower temperatures within the high-temperature regime $k_B T \gg \hbar\omega_{ZB}$, and corresponds to similar LSSE studies in YIG[30], [45], [54]. Further, Figure 2(c) shows that the SSR monotonically increases with thickness and shows no indications of saturation, indicating that the magnon diffusion length is at least 1 μm at room temperature, similar to the magnon diffusion length in single-crystal YIG of 1-10 μm[29], [55]. The temperature dependent measurements also show that the magnetic hysteresis becomes softer at lower temperatures (see Supplement), consistent with temperature-induced strain-dependent magnetic anisotropy fields in V[TCNE]$_x$[42], [56].

These experimental results are compared to the theoretical model using the temperature-dependence of the magnon scattering rates[42] and saturation magnetization[46] for V[TCNE]$_x$. The SSE voltage is calculated as a function of temperature using Eq. 2 for the V[TCNE]$_x$ thicknesses used in the experiment, as shown in Fig. 3(a). A single temperature-dependent $\tau_m$ is used for all the curves. Therefore, the decrease of the experimental V$_{SSE}$ with increasing temperature (for $T \gtrsim 150K$) is attributed to the shortening of the relaxation time of magnon-conserving processes, consistent with prior studies of the thermal conductivity of magnons for a variety of materials, including YIG[57]–[59]. The dashed lines accompanying the solid theoretical results in Fig. 3(a) account for a 10%-error on the thickness size estimate[60]. It is also interesting to note that the $V_{SSE}$ values reported here are comparable to the ones measured in YIG[7], [29], [45], [61], despite the larger exchange parameters and saturation magnetization in YIG, which lead to a larger group velocity and energy band width (larger number of magnons). However, as noted above, the spin current pumped into the metal is proportional to $1/M_S$ as can be seen in the expression for the spin pumping current, $j_s^{pumping} = \frac{g_r^{\uparrow\downarrow}}{4\pi M_S^2} \boldsymbol{M} \times \frac{\partial \boldsymbol{M}}{\partial t}$, where $\boldsymbol{M} = (m_x, M_S, m_z)$ is the total magnetization[33], [34]. This dependence therefore *suppresses* spin injection for YIG while *enhancing* spin injection for V[TCNE]$_x$.

Using the same set of fit parameters extracted from the analysis above, in Fig. 3(b) $V_{SSE}$ is plotted as a function of the V[TCNE]$_x$ thickness $d$ for different temperatures, and good agreement between theory and experiment is seen without adjusting parameters other than the thickness. For the measured experimental temperature range and the thickness values, a magnon diffusion length $\lambda_m \approx 1 - 2$ μm is obtained for V[TCNE]$_x$, shown in Fig. 3(c). This is consistent with the linear behavior of the $V_{SSE}$ with respect to $d$, shown in Fig. 3(b).

With this validation of the ability of the extended theory presented here to predict this new regime of high efficiency SSE, other magnonic transport parameters can now be determined. Specifically, the diffusion coefficient for V[TCNE]$_x$, $D_m = \mu k_B T$, where $\mu$ is the magnon mobility, is related to the magnon diffusion length via $\lambda_m = \sqrt{D_m \tau_{th}}$. $D_m$ is plotted in Fig. 3(d), with $D_m = 0.002 - 0.01$ cm²/s for the temperature range analyzed in this work. This diffusion coefficient is several orders of magnitude smaller than that of YIG[7], [29], which is a consequence of the smaller V[TCNE]$_x$ exchange constant ($D_{ex}$) as $D_m \sim D_{ex}^2$ (see Supplement). However, since

$\lambda_m$ of V[TCNE]$_x$ and single-crystal YIG are similar at around 1 $\mu$m at room temperature, as discussed above, it appears that magnon lifetimes, $\tau_{th}$, in V[TCNE]$_x$ must exceed those in YIG by at least one order of magnitude in order to account for this smaller diffusion coefficient. Further, as discussed above, the magnetic ordering temperature is dependent on the exchange stiffness. The remarkably small exchange stiffness in V[TCNE]$_x$ suggests an ordering temperature around or below 10 K (see Supplement). However, V[TCNE]$_x$ thin films and bulk samples have demonstrated ordering temperatures exceeding 600 K[41], [62] and the devices here demonstrate efficient spin pumping at room temperature. This combination of small exchange $D_{ex}$ ($J$) and high $T_C$ is not well understood, highlighting the opportunity for future work in understanding magnetic ordering in this class of molecule-based magnets and developing a new class of SSE-active materials.

In summary, our theoretical model provides a robust platform to investigate alternative materials that may produce large SSE voltages. Importantly, as the injected spin current into the heavy metal layer is inversely proportional to the saturation magnetization $M_s$, materials with low $M_s$ are ideal candidates for such applications. Further, the injected spin current is also proportional to the thermal gradient in the material, suggesting low thermal conductivity materials (such as V[TCNE]$_x$ and YIG – see Supplement) are ideal as they will produce larger thermal gradients given the same input heat flux. In general, materials with long magnon lifetimes, long magnon scattering times, small $M_s$, and large exchange values are optimal for such applications.

**DISCUSSION**

We have presented an extended theoretical model for the SSE wherein we build on previous theory describing the bulk magnon spin current in a FM layer by accounting for the temperature- and thickness-dependence of magnon non-conserving processes. This theory quantitatively describes prior SSE experiments on YIG, providing a physical explanation for the features seen in the temperature- and thickness-dependencies. Within this context, we identify optimized magnetic material parameters for efficient spin pumping, including low-$M_S$ (small magnon bandwidth) and long magnon scattering time. To further validate this theory, we provide systematic measurements of the LSSE in the low-$M_S$, low-damping, and low-bandwidth ferrimagnet V[TCNE]$_x$ for various thicknesses and temperatures, providing direct confirmation and excellent theory-experiment agreement. These studies provide a foundation for future research and applications of molecule-based magnetic materials in spintronics and spin caloritronics, and has the potential to impact developments in QISE studies and applications that depend on V[TCNE]$_x$[63], [64]. Notably, this work highlights the potential of ultra-low $M_s$ materials for efficient spin injection and spin pumping applications.

## ACKNOWLEDGEMENTS

S. W. K. and D. R. C. wrote the manuscript. S. W. K. deposited V[TCNE]$_x$ films and encapsulated with glass layer. Y. Z. and B. W. mounted heating elements and carried out SSE measurements. S. W. K. analyzed and extracted values from the experimental data. S. W. K., D. R. C., M.E.F, and E. J.-H. discussed the extended theory, and D. R. C. carried out analytical calculations. M. J. N and S. C. deposited Pt films. E. J.-H. and J. P. H. conceived the experimental idea. All authors discussed the results and revised the manuscript. S. W. K. and E. J.-H. acknowledge funding from NSF DMR-1808704. Y. Z. and J. P. H. acknowledge the NSF MRSEC "Center for Emergent Materials" (CEM) at The Ohio State University funded by NSF DMR-2011876. B. W. acknowledges funding from the US Army Research Office (ARO) grant W911NF2120089. D. R. C. and M. E. F. acknowledge funding from NSF DMR-1808742. M. J. N. and R. K. K.

acknowledge funding from DAGSI RX14-OSU-19-1, and S. C. and R. K. K. acknowledge funding from the DARPA D18AP00008.

# Figures and Captions

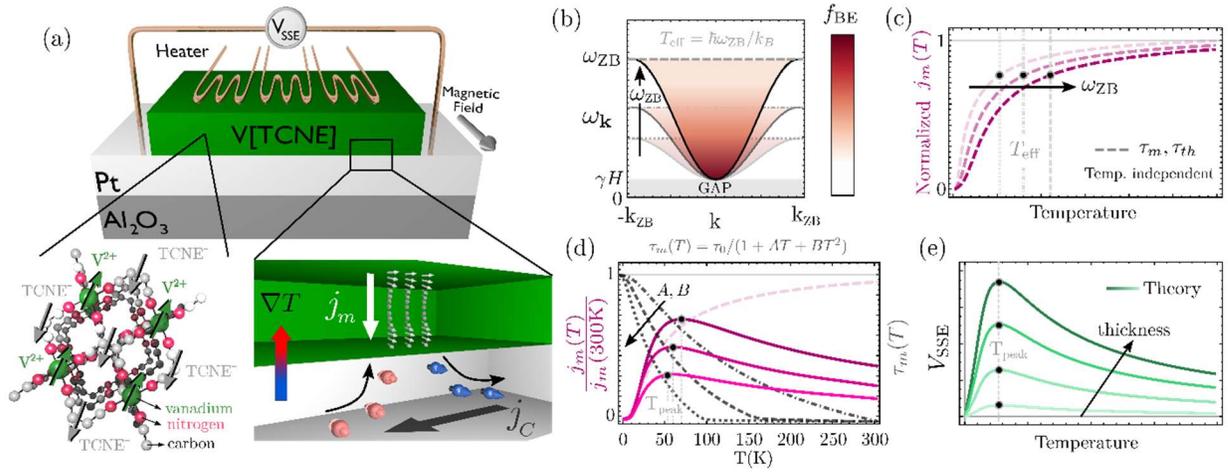

Figure 1: (a) Schematic of the conventional setup for measuring the longitudinal SSE voltage $V_{SSE}$ in the V[TCNE]$_x$/Pt/Al$_2$O$_3$ sample. The structure of V[TCNE]$_x$ is displayed on the bottom left panel, and the spin-charge current conversion due to ISHE is shown in the bottom right panel. (b) V[TCNE] acoustic magnon energies as a function of wavevector for three different values of $\omega_{ZB}$. The density plot represents schematically the magnon population as a function of the magnon frequency. (c) Normalized spin current $j_m$ [Eq. (1)] at the V[TCNE]$_x$/Pt interface as a function of temperature for the corresponding three different values of $\omega_{ZB}$, and assuming temperature-independent $\tau_m, \tau_{th}$. For increasing $\omega_{ZB}$ (following the arrow), the spin current requires higher temperatures to saturate and results in higher $T_{eff}$ temperatures. (d) Normalized $j_m$ spin current (left axis) taking into account the temperature dependence of the magnon scattering time $\tau_m$ (right axis). As the magnon scattering time decreases (i.e. the parameters $A, B$ increase – following the arrow – which correspond to magnon-conserving relaxation rates due to temperature-dependent magnon scattering processes), the peak temperature $T_{peak}$ position shifts to lower temperatures. (e) $V_{SSE}$ [Eq. (2)] vs temperature for different FM thicknesses. As the thickness of the FM layer increases (following the arrow), the magnitude of the injected spin current (hence the measured ISHE voltage) also increases until the signal saturates as the film thickness exceeds the magnon diffusion length.

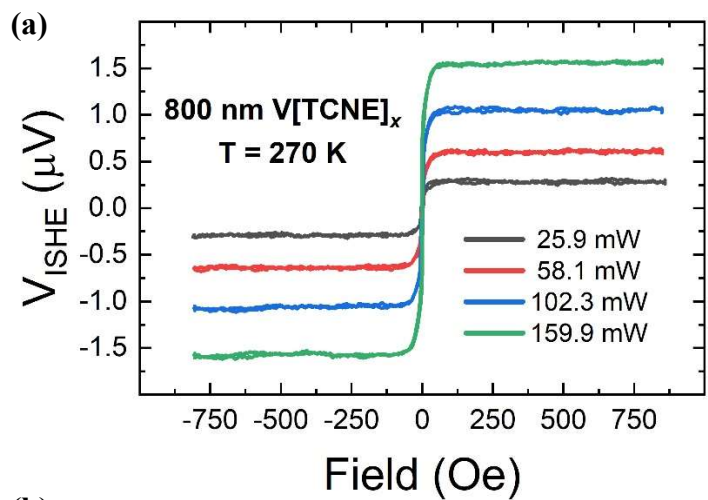
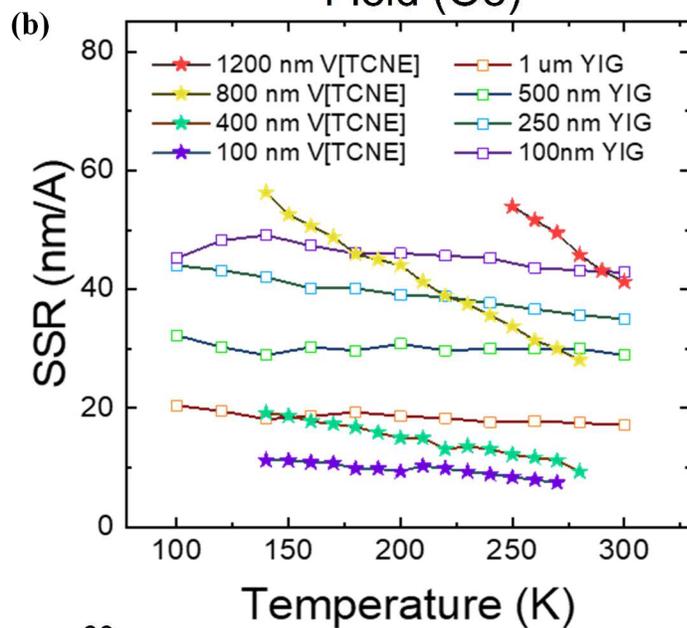
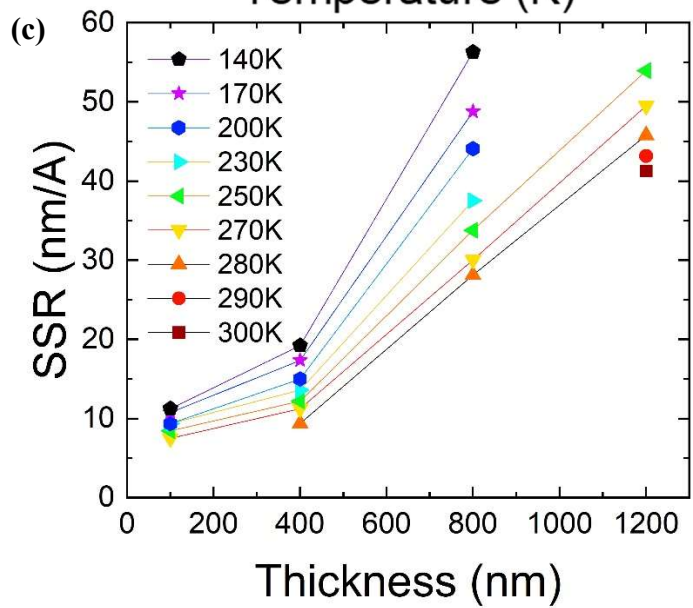

Figure 2: (a) Raw data of ISHE voltage versus applied field with various heater powers measured in the Pt layer from an 800 nm thick V[TCNE]$_x$ sample. (b) SSR versus temperature for various thicknesses of V[TCNE]$_x$ (filled stars) and YIG (empty squares – from Prakash *et al.* – Ref. 45) (c) Select V[TCNE]$_x$ data from (b) plotted versus V[TCNE]$_x$ film thickness.

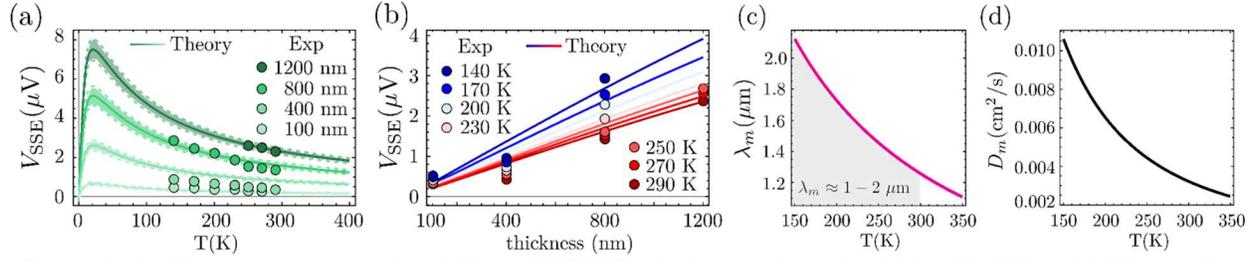

Figure 3: (a) $V_{SSE}$ vs temperature for different V[TCNE] thickness values $d$ = 100, 400, 800 and 1200 nm. Green circles represent the experimental data while solid lines represent the plot of the theoretical modeling using Eq. (2). The dashed solid lines accounts for a ±10% error on the thickness values. (b) $V_{SSE}$ vs thickness size for different temperature. The circles represent the experimental data while the solid lines the theoretical predictions. (c) Plot of extracted magnon diffusion length $\lambda_m$ of V[TCNE]$_x$ as a function of temperature. (d) Plot of the extracted magnon diffusion coefficient $D_m$ for V[TCNE]$_x$ as a function of temperature.

# Supplemental Material

**Magnon Diffusion Length and Longitudinal Spin Seebeck Effect in Vanadium Tetracyanoethylene (V[TCNE]$_x$, $x \sim 2$)**

Seth W. Kurfman[1*], Denis R. Candido[2*], Brandi Wooten[3], Yuanhua Zheng[4], Michael J. Newburger[1], Shuyu Cheng[1], Roland K. Kawakami[1], Joseph P. Heremans[1,3,4], Michael E. Flatté[2,5], and Ezekiel Johnston-Halperin[1]

[1]Department of Physics, The Ohio State University, Columbus, OH 43220, USA

[2]Department of Physics and Astronomy, The University of Iowa, Iowa City, IA 52242, USA

[3]Department of Materials Science and Engineering, The Ohio State University, Columbus, OH 43220, USA

[4]Department of Mechanical Engineering, The Ohio State University, Columbus, OH 43220, USA

[5]Department of Applied Physics, Eindhoven University of Technology, Eindhoven, The Netherlands

[*]These authors contributed equally

## 1. Device Fabrication and Measurement

For all devices, 7.5 nm of Pt is deposited on 430 μm sapphire C-plane substrates (3 x 8 mm$^2$) via molecular beam epitaxy (MBE) and transferred under ultra-high vacuum (UHV) into the V[TCNE]$_x$ growth glovebox (argon atmosphere with $O_2$, $H_2O$ < 0.1 ppm). This UHV transfer limits surface contamination of the Pt by $O_2$ and other gases, $H_2O$, or other particulates that may interfere with spin injection across the V[TCNE]$_x$/Pt interface. V[TCNE]$_x$ thin films are then deposited on the Pt in a ~3 x 3 mm$^2$ area and subsequently encapsulated with an OLED epoxy (Ossila E151) and 100 μm-thick glass cap to protect the films from oxygen-, moisture-, and solvent-exposure risks present through the remaining steps of device fabrication and measurement. On top of the glass cap, a Cu heat spreader and resistive heater is mounted with GE Varnish to ensure good thermal contact. The device is loaded into a cryostat (Lakeshore ST-300) for measurements such that the external magnetic field is in the plane of the magnetic film. A type-T thermocouple was placed below the device to read the temperature of the sample during measurements. Cu wires (adhered to the Pt layer by indium pressing) were used as leads to record the voltages, obtained through a Keithley nanovoltmeter model 2182a. All device fabrication steps took place inside glovebox environments to protect the V[TCNE]$_x$ from exposure to oxygen and moisture. The cryostat was sealed in the glovebox and immediately connected to vacuum to conduct the measurements.

## 2. Longitudinal Spin Seebeck effect

The derivation of the equations describing the longitudinal spin Seebeck effect will be done here similarly to the theory developed in Refs. [1, 2]. We consider this theory as time-resolved studies have shown the temperature-dependent contribution to the LSSE is due to the bulk magnon spin current, while the temperature-independent contribution is due to the interfacial temperature difference between the magnons in the ferromagnet and the electrons in the heavy metal[36]. Only small modifications will be performed in order to address the particular characteristics of our material V[TCNE]$_x$, along with additions towards improving of the theory to account for signatures seen in prior SSE experiments.

## 2.1 Bulk magnon spin current

As our magnetic material is polarized along the y direction, the spin current tensor due to magnons is given by

$$j_m^y = \frac{\hbar}{(2\pi)^3} \int d\mathbf{k}\, \mathbf{v}(\mathbf{k})\, n_k, \quad (S1)$$

where $\hbar$ is the reduced Planck's constant, $\mathbf{v}(\mathbf{k}) = \left(\frac{1}{\hbar}\right) \nabla_k (\hbar\omega_k)$ is the velocity of magnons, $n_k = 1/[e^{(\hbar\omega_k - \mu)/k_B T} - 1]$ is the number of magnons with energy $\hbar\omega_k$ and wavevector $\mathbf{k}$, $T$ is the temperature, $k_B$ is the Boltzmann constant and μ is the chemical potential associated to the magnons. Due to the small saturation magnetization of V[TCNE]$_x$, $4\pi M_S \approx 70 - 95\ G$[46], [47], [65], small applied magnetic field, and the range of temperatures of our experiment (150 – 300 K), the dispersion of the relevant magnons conducting the spin current can be well approximated by a nearly parabolic dispersion, namely,

$$\omega_k = \gamma \sqrt{(H + D_{ex}k^2)\left(H + M_0 \left(\frac{k_x^2 + k_z^2}{k^2}\right) + D_{ex}k^2\right)}, \quad (S2)$$

$$\approx \gamma H + \gamma D_{ex} k^2, \quad (S3)$$

where H is the applied magnetic field, $\gamma/2\pi = 2.73 \times 10^6\ s^{-1}/Oe$ [65] is the gyromagnetic ratio and $D_{ex} = M_S \lambda_{ex} = 5.82 \times 10^{-11}\ Oe\ cm^2$ is the exchange constant (or spin wave stiffness) for $4\pi M_S = 95\ G$ [46], $\lambda_{ex} = \frac{2A_{ex}}{\mu_0 M_S^2} = 6.13 \times 10^{-1}\ m^2$ and $A_{ex} = 2.2 \times 10^{-1}\ J/m$ [65]. Although different values for V[TCNE]$_x$ $4\pi M_S$ were already reported[46], [65], only one work investigated the exchange constant[65]. As the $\lambda_{ex}$ depends strongly on $4\pi M_S$, we expect that different samples (with different $4\pi M_S$) contains different $D_{ex}$. Here, we have used $D_{ex} = 1.74 \times 10^{-10}\ Oe\ cm^2$. It is well known that this dispersion relation Eq. (S3) will not produce realistic transport results for $k_B T \gg \hbar \omega_k$. This happens because the magnon energy dispersion in Eq. (S3) is actually periodic in $k$ and does increase monotonically as a function of $k$. Accordingly, we have to replace the low-frequency dispersion by lattice results, which corresponds to[7]

$$\omega_k = \gamma H + \omega_{ZB}\left(1 - \cos\frac{\pi k}{k_{ZB}}\right), \quad (S4)$$

with $\omega_{ZB} = \frac{8\gamma}{\pi^2} D_{ex} k_{ZB}^2$, $k_{ZB} = 2\sqrt{3}\pi/a$, and $a$ being the averaged distance between two unit cells (here we have assumed spherical BZ with $a \approx 7.35\ \text{Å}$, defined by the spatial distance between vanadium positions).

If the system is in an equilibrium state, $n_k = n_k^{eq}$ is constant across our sample and $j_m^y = 0$ follows from Eq. (S1). Accordingly, a finite current can only occur when the magnons are out of equilibrium. This can happen, for instance, when there is a spatial dependence on $n_k$ which can occur when there is a spatial dependence on temperature. If we denote the out of equilibrium distribution as $n_k(r)$, the generated current will read

$$j_m^y = \frac{\hbar}{(2\pi)^3}\int d\mathbf{k}\, v_k\left[n_k(\mathbf{r}) - n_k^{eq}\right]. \quad (S5)$$

The number of magnons in the presence of a spatially dependent temperature along $z$ can be determined by means of Boltzmann equation assuming the relaxation-time approximation (similarly to Ref. [7]), and reads

$$n_k(z) = n_k^{eq} + \frac{\partial n_k^{eq}}{\partial(\hbar\omega_k)}\left\{-\mu(z) - \frac{(\hbar\omega_k)}{T}[T(z) - T] + \sum_{i=0}^{\infty} g^i(z)P_i(\cos\theta)\right\}. \quad (S6)$$

For the spin current calculations, the only term that contributes to a non-vanishing spin current is the one proportional to $g^1(z)$,

$$g^1(z) \approx -\tau_m \frac{\mathcal{I}_2}{\mathcal{I}_1}\left[\frac{\partial\mu(z)}{\partial z} + \frac{\hbar\omega_k}{T}\frac{\partial T(z)}{\partial z}\right], \quad (S7)$$

$$\mathcal{I}_n = \int \frac{d\mathbf{k}}{(2\pi)^3} |v_k|^n \frac{\partial n_k^{eq}}{\partial(\hbar\omega_k)}, \quad (S8)$$

with $\tau_m$ being the relaxation time of process conserving the number of magnons (e.g., magnon scattering by a paramagnetic impurity, two-magnons and four-magnons scattering processes). By projecting the incident spin current along the normal vector of the interface $z = d$, we obtain

$$j_m^y = \boldsymbol{j}_m^y \cdot \hat{z} = j_m^{\nabla T} + j_m^{\nabla\mu}, \quad (S9)$$

$$j_m^{\nabla T} = -\hbar \frac{\tau_m}{3}\int \frac{d\mathbf{k}}{(2\pi)^3}\frac{\partial n_k^{eq}}{\partial(\hbar\omega_k)}|v_k|^2 \frac{\hbar\omega_k}{T}\frac{\partial T}{\partial z} = -C_z\frac{\partial T}{\partial z}, \quad (S10)$$

$$j_m^{\nabla\mu} = -\hbar\frac{\tau_m}{3}\int \frac{d\mathbf{k}}{(2\pi)^3}\frac{\partial n_k^{eq}}{\partial(\hbar\omega_k)}|v_k|^2 \frac{\partial\mu}{\partial z} = \hbar\frac{\tau_m \mathcal{I}_2}{3}\frac{\partial\mu}{\partial z}. \quad (S11)$$

Therefore, we can see that a bulk magnon spin current is produced by both gradient of temperature and magnon accumulation (gradient of chemical potential). Additionally, it is known that the chemical potential obeys the diffusion equation[66]

$$\frac{\partial^2 \mu(z)}{\partial z^2} = \frac{1}{\lambda_m^2} \mu(z), \qquad (S12)$$

with magnon diffusion length $\lambda_m = \sqrt{D_m \tau_{th}}$, $D_m = \tau_m \mathcal{J}_2/(3\mathcal{J}_0)$, with $\tau_{th}$ being the relaxation time of processes not conserving the number of magnons (i.e. magnon lifetime). Accordingly, the general solution for the chemical potential reads

$$\mu(z) = c_1 e^{z/\lambda_m} + c_2 e^{-z/\lambda_m}, \qquad (S13)$$

and the incident spin current follows immediately as

$$j_m^y(z) = -C_z \frac{\partial T}{\partial z} + \frac{\hbar D_m}{\lambda_m} \left[ c_1 e^{\frac{z}{\lambda_m}} + c_2 e^{-\frac{z}{\lambda_m}} \right]. \qquad (S14)$$

The constants $c_1$ and $c_2$ are determined by the boundary conditions of the problem. The first is obtained such that there is no current at $z = 0$, while the second one is determined by knowing that the spin current pumped across the interface is given by[33], [34]

$$\lim_{z \to d^+} j_m^{pumping}(z) = \frac{\hbar g_r^{\uparrow\downarrow}}{4\pi M_S^2} \boldsymbol{M}(\boldsymbol{r},t) \times \left.\frac{\partial \boldsymbol{M}(\boldsymbol{r},t)}{\partial t}\right|_{z=d}, \qquad (S15)$$

where $M_S$ is the saturation magnetization, $\boldsymbol{M}(\boldsymbol{r},t)$ is the total magnetization and $g_r^{\uparrow\downarrow}$ the real part of the spin pumping conductance, which for V[TCNE], $g_r^{\uparrow\downarrow} = 0.95 \times 10^{18}\ m^{-2}$ [67]. Using the corresponding boundary conditions, we obtain the final expression for the spin current at the interface, namely

$$j_m^y(z = d) = -F \frac{B_1 B_S}{\sqrt{B_0 B_2}} \tanh\left(\frac{d}{2\lambda_m}\right) g_r^{\uparrow\downarrow} \frac{\partial T}{\partial z}, \qquad (S16)$$

with

$$F(T) = \frac{\gamma \hbar k_B k_{ZB}^2 \omega_{ZB} \sqrt{\tau_{th}\tau_0}}{2\pi\sqrt{3} \times 4\pi M_S}, \quad (S17)$$

$$B_S = \int_0^1 dq\, q^2 \sin^2\left(\frac{\pi}{2}q\right) \frac{xe^x}{\eta_q(e^x - 1)^2}, \quad (S18)$$

$$B_0 = \int_0^1 dq\, q^2 \frac{x}{e^x - 1}, \quad (S19)$$

$$B_1 = \int_0^1 dq\, q^2 \frac{x^2}{e^x - 1}, \quad (S20)$$

$$B_2 = \int_0^1 dq\, q^2 \sin^2\left(\frac{\pi}{2}q\right) \frac{x}{\eta_q(e^x - 1)^2}, \quad (S21)$$

$q = k/k_{ZB}$, $x = \hbar\omega_k/k_B T$, and $\tau_m = \tau_0/\eta_q$ with $\tau_0$ being the relaxation time for magnons with $\boldsymbol{k} = 0$ and $\eta_q$ being the wavevector dependence of the relaxation time, which will be commented on in section **2.3** below. Once the spin current $j_m^y$ enters the metal region, its magnitude begins to diminish. This happens due to the spin diffusion and relaxation mechanisms that take place within the metal region[34]. As a consequence, this gives rise to a characteristic diffusion length given by $\lambda_N$ and corresponding spin current[11], [34]

$$j_m^y(z) = j_m^y(z = d) \sinh\left(\frac{d + t_{Pt} - z}{\lambda_N}\right) / \sinh\left(\frac{t_{Pt}}{\lambda_N}\right), \quad \text{for } d < z < d + t_{Pt}. \quad (S22)$$

## 2.2 Inverse Spin Hall Effect

Once we have an incident spin current $\boldsymbol{j}_m^\sigma$ in the metal, this is transformed to a perpendicular charge current $\boldsymbol{j}_C$ via the inverse spin Hall effect[35]. The expression that relates these two currents is given by

$$j_C(z) = \theta_{SH}\left(\frac{2e}{\hbar}\right) j_m^\sigma(z) \times \boldsymbol{\sigma}, \qquad (S23)$$

where $\theta_{SH}$ is the spin-Hall angle, $e$ the elementary charge, and $\boldsymbol{\sigma}$ is the spin polarization. For our case, since the magnetization of the material is along $y$ and the temperature gradient along $z$, we have $\boldsymbol{\sigma} = \hat{y}$ and $\boldsymbol{j}_m^\sigma \propto j_m^y \hat{z}$, yielding

$$j_C(z) = -\theta_{SH}\left(\frac{2e}{\hbar}\right) j_m^y \hat{z} \times \hat{y},$$

$$j_C(z) = -\theta_{SH}\left(\frac{2e}{\hbar}\right) j_m^y \hat{x}. \qquad (S24)$$

The electric current flowing within the normal metal produces a voltage across $x$ given by $V_{SSE} = I_C R_{Pt}$, with $R_{Pt}$ being the resistance of the Pt metal, and a total current given by

$$I_C = \int_0^{l_w} dy \int_d^{d+t_{Pt}} dz\, j_m^y(z),$$

$$= l_w \lambda_N \tanh\left(\frac{t_{Pt}}{2\lambda_N}\right) j_m^y(z=d). \qquad (S25)$$

Accordingly, the spin Seebeck voltage reads

$$V_{SSE} = R_{Pt} l_w \lambda_N \theta_{SH}\left(\frac{2e}{\hbar}\right) \tanh\left(\frac{t_{Pt}}{2\lambda_N}\right) j_m^y(z=d). \qquad (S26)$$

Although we have not measured the Pt resistance, $R_{Pt}$, the same can be estimated assuming the Pt strip dimensions $(l, l_w, t_{Pt})$ and conductivity $\sigma_{Pt} = 2.4 \times 10^4\ \Omega^{-1}\ cm^{-1}$,

$$R_{Pt} = \frac{1}{\sigma_{Pt}} \times \frac{length}{cross - area} = \frac{1}{\sigma_{Pt}} \frac{l}{t_{Pt} l_w}. \qquad (S27)$$

### 2.3 Magnon relaxation scattering times

Usually, the voltage produced by the longitudinal spin Seebeck effect is measured for different temperatures. Within the equations above, many material parameters contain a temperature dependence, e.g., $M_S, \tau_m$ and $\tau_{th}$. Although the temperature dependence of the relaxation time of processes conserving the number of magnons ($\tau_m$) was included in the original papers[7], [29], the temperature dependence of $\tau_{th}$ was not. We emphasize, however, that in reality, $\tau_{th}$ possesses a strong temperature dependence not only in YIG material[40] but also in V[TCNE]$_x$[42]. Accordingly, the temperature dependence of $\tau_{th}$ will also be included in the SSE calculation presented in this work. Previous work on the V[TCNE]$_x$ magnon lifetime[42] has shown that the magnon scattering process that does not conserve the number of magnons are predominately given by the exchange interaction ($\hbar\omega_{int}$) between magnons and paramagnetic defect spins (also known as two-level spin fluctuators). Accordingly, we have $\tau_{th} = 1/\gamma\Delta H$ with $\Delta H$ being the magnon linewidth accounting for a line shape factor for the finite two-level spins $\frac{1/t_s}{\left(\hbar^2/t_s^2 + (\hbar\omega_{k=0} - \hbar\omega_{eg})^2\right)}$, and the population ratio between ground and excited impurity states. This results in[40], [42]

$$\Delta H = \Delta H_0 + \frac{S}{\gamma}\frac{N_{imp}}{N}(\hbar\omega_{int})^2 \frac{1/t_s}{\left(\hbar^2/t_s^2 + (\hbar\omega_{k=0} - \hbar\omega_{eg})^2\right)}\tanh\left[\frac{\hbar\omega_k}{2k_B T}\right], \quad (S28)$$

where $N_{imp}/N$ is the ratio between number of impurities and number of V[TCNE] atoms, $S$ is the average V[TCNE]$_x$ spin per site, and $\Delta H_0$ is a linewidth due to other relaxation mechanisms. Here, we will assume two-level spin lifetime $t_s = t_\infty e^{E_b/k_B T}$ with $t_\infty$ being the two-level spin lifetime at infinite temperature, and $E_b$ a phenomenological activation energy[40]. Although the formula above was derived for uniform magnon excitations, i.e., $\boldsymbol{k} = 0$, we assume a generalization can be done for $\boldsymbol{k} \neq 0$. In the simulations we will be using $S \approx 1, \omega_{int} \approx \omega_{eg}, E_b = 1\,meV$, and $\omega_{eg} N_{imp}/N = 36.5\,GHz$[42]. As for the relaxation time of processes conserving the number of magnons, $\tau_m$, there is still no developed theory for the V[TCNE]$_x$ material. Accordingly, here we will be using expressions obtained and developed for the YIG material. As both YIG and V[TCNE]$_x$ are ferrimagnetic, we will use the same algebraic expression for $\tau_k$, namely[7]

$$\tau_k = \frac{\tau_0}{\eta_q}, \qquad (S29)$$

$$\eta_q = 1 + A_1 qT + A_2 q^2 T^2 + A_3 q^3 T^2. \, (S30)$$

The details and particularities of the V[TCNE]$_x$ material enters within the constants $A_1$, $A_2$ and $A_3$. Accordingly, in this work, these are going to be our fitting parameters. They are adjusted so that we produce the best fit to the experimental data of Fig. 3 in the main manuscript. We stress that our fitting parameters are the same for all data sets corresponding to different thicknesses. In this work, the best fits were obtained by $A_1 = 25.00$, $A_2 = 0.84$ and $A_3 = -0.54$.

As we can see from Eq. (S26), the only *explicit* thickness-dependence of V$_{SSE}$ appears through the $\tanh(d/2\lambda_m)$ factor taking into account the diffusion of magnons. Accordingly, SSE measurements of samples with different thicknesses will only be different by an overall factor, yielding a maximum peak position of V$_{SSE}$ as a function of temperature that is independent of the thickness values. We emphasize that this is a rather simple picture, as it ignores, for example, the thickness-dependence of the magnetic properties of the material. For instance, SSE measurements in YIG show a strong thickness-dependence of the maximum peak of V$_{SSE}$ vs temperature[30], [44] – while this peak feature is generally present in the current theory there remain several features not well explained[7], [29]. Additionally, it can also be seen that at room temperature, some samples with different thicknesses produce similar V$_{SSE}$[30], corroborating some thickness-independent SSE which requires further investigation. Accordingly, a proper assessment of the SSE in magnetic materials also needs to consider the thickness dependencies within the magnetic material properties. Here, we improve the theory developed by Rezende[7], [29] by including the thickness dependence of the magnon scattering times. We show this improvement provides a clearer and more accurate prediction of the main features of the $V_{SSE}$ illustrated in Fig. S1. It has been known for many years that a great contribution of the YIG magnon linewidth is due to the magnon scattering at surface imperfections (or pits)[40], [43]. Accordingly, smaller bulk samples result in shorter magnon scattering times. These processes were calculated[40], [43] assuming scattering centers at the surface coupled to magnons via dipole-dipole interaction, yielding for a single scattering center

$$\frac{1}{\tau_d} = M_s \frac{V_{pit}}{V}, \qquad (S31)$$

where $V_{pit}/V$ is the relative volume of the pit with respect to the sample volume. This dependence was experimentally realized by noting that the magnon linewidth $\Delta H = \gamma \tau_d$ was proportional to the size of the polishing spheres used for polishing the YIG surface[40], [43]. Accordingly, if we assume different sample sizes with the same area and same number of pits, but different thicknesses, the thickness ($d$) dependence of this relaxation process becomes

$$\frac{1}{\tau_d} = M_s \frac{V_{pit}}{d \times Area}, \qquad (S32)$$

with $d$ being the sample thickness.

Therefore, the magnon scattering time, taking into account the thickness, wavevector, and temperature dependences, reads

$$\frac{1}{\tau(\mathbf{k}, d)} = \frac{1}{\tau_\mathbf{k}} + \frac{1}{\tau_d} = \frac{\eta_{q,d}}{\tau_0}, \qquad (S33)$$

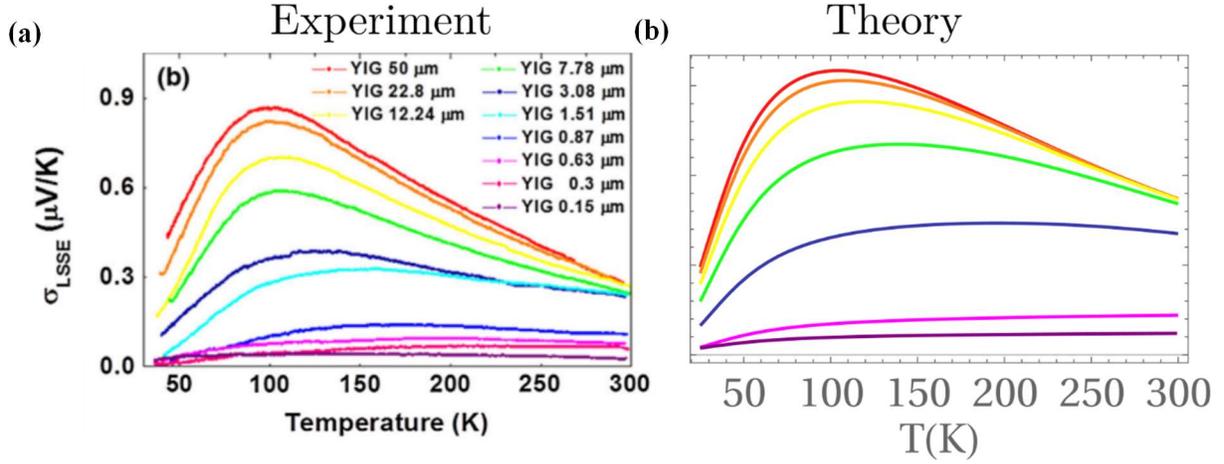

**S4:** (a) Experimental data of SSE in YIG films of various thicknesses vs temperature. As the thickness of the YIG film increases, the temperature peak position of the SSE signal decreases. Adapted from Guo *et al*. [Ref. 15]. (b) Modeled temperature dependence behavior of the peak position of $V_{SSE}$ vs temperature for different YIG film thicknesses, corroborating the signatures seen in prior experimental studies in YIG seen in Refs. 15 and 16.

with $\eta_{q,d} = \eta_q + \tau_0/\tau_d$. It is interesting to note that since this scattering mechanism is mediated via dipole interaction, the smaller saturation magnetization of V[TCNE]$_x$ (compared to YIG) will benefit from the suppression of this process. As we show in Fig. S1, the inclusion of the thickness dependence is able to explain very well all the features of the $V_{SSE}$ vs temperature in YIG. When the temperature increases, the magnon diffusion length diminishes due to the decrease of the magnon lifetime. For temperatures such that $d \ll 2\lambda_m$ and $\tanh(d/2\lambda_m) \to 1$, we obtain the absence of the thickness dependence on the signal of SSE. As we already commented above, the peak of $V_{SSE}$ vs temperature happens due to an interplay between the unbalance of the excited number of magnons carrying the spin current, which increases with temperature, and their scattering times, which decreases with temperature. Due to the smaller exchange interaction, the temperature peak for V[TCNE]$_x$ is expected to occur at small temperatures ($T \approx 20\ K$). Since the experimental data for the temperature dependence of $V_{SSE}$ is far away from this region, we do not include the thickness dependence on the magnon scattering time as at high temperature the dominant relaxation mechanism is $\tau_k$ [Eq. S29]. It was also found in previous work[30] that the peak position of the SSE signal with temperature is impacted by the metal layer choices and interfaces, as well as the strength of the applied magnetic field. The same description here may account for these changes as well to some extent, as in general the peak position moves to lower temperatures as the magnon scattering times and lifetimes increase. It should be noted also that despite this feature, as V[TCNE]$_x$ is a ferrimagnetic material, one also expects higher energy magnon bands (optical magnon branches) with opposite angular momentum, similar to what has been seen in the spectrum of YIG[68]. Accordingly, once the temperatures are high enough to excite these higher magnon bands, a corresponding suppression of the SSE voltage is expected. This has already been seen for different ferrimagnets[17] and antiferromagnetic materials[23], [24]. However, due to the lack of information about the higher V[TCNE]$_x$ magnon bands, the corresponding calculation has not been performed. Despite this, one cannot exclude the excitation of these higher magnon bands (with opposite polarization) as an additional reason underlying the suppression of the V$_{SSE}$ for temperatures $T \gtrsim 150$ K.

When considering a quadratic dispersion of exchange spin waves, where $E(k) = Dk^2 = \left(\frac{\hbar^2}{2m}\right)k^2$, the small spin wave stiffness $D_{ex}$ of V[TCNE]$_x$ (two orders of magnitude smaller than YIG) suggests the effective magnon masses in V[TCNE]$_x$ are particularly large compared to typical magnetic materials. This aligns well with the small diffusion coefficient of V[TCNE]$_x$ $D_m$, as this suggests a small drift velocity consistent with classical particles having large mass impacted by an external force. Additionally, in essence, a small exchange constant results in a narrow magnon energy bandwidth, which implies a smaller number of magnons diffusing at a specific temperature – however, the magnon density of states increases with decreasing spin wave stiffness hence the narrow bandwidth increases the magnon density for a given frequency spread. Further, the thermal magnon de Broglie wavelength is dependent on the spin wave stiffness $D_{ex}$ via $\lambda_{th} \propto \sqrt{D_{ex}}$ [12], [69] yielding $\lambda_{th_{V[TCNE]}} \sim 72$ pm which is approximately the radius of C and N atoms and well below the radius of a V atom where unpaired spins are predominantly located. This potentially provides insight and an explanation to the surprisingly long magnon lifetimes, where the small effective magnon scattering cross-section reduces scattering from spins and impurities. Additionally, as the magnon contribution to the thermal conductance $\kappa \propto D_{ex}$,[12] magnons likely have little contribution to the thermal conductance of V[TCNE]$_x$ though further studies are required to explore this contribution.

In terms of the magnon lifetime exceeding 1 µs, a potential explanation for this surprisingly robust magnon lifetime can be found in a consideration of the magnon-scattering processes, including 2-, 3-, and 4-magnon scattering, that are proportional to the magnetization, uniformity, and band structure of the magnetic material. Due to the low-$M_S$ and small $D_{ex}$ of V[TCNE]$_x$, the narrow magnon bandwidth suppresses 3-magnon scattering by narrowing the phase space for frequency-conserving scattering processes (i.e. $\omega_1 = \omega_2 + \omega_3$). Furthermore, magnon impurity scattering proceeds via exchange ($\sim D_{ex}$) and dipole interactions ($\sim M_S$), both of which are small in V[TCNE]$_x$ hence resulting in long scattering times and magnon lifetimes. This likely has an impact on the extracted magnon diffusion length in V[TCNE]$_x$ within the main text over 1 µm at room temperature. As V[TCNE]$_x$ lacks long-range structural order[47], this micron length scale magnon diffusion length is unprecedented in a disordered material and significantly exceeds the 100 nm magnon diffusion length in polycrystalline YIG films[31].

We present a physical description for these signatures and account for this temperature-dependence by considering the impact of film thickness on the magnon scattering time which has not been included in prior SSE theory. That is, for thinner films, surface scattering affects the scattering rates more than in thick films so that thicker films should experience smaller scattering rates. Our theory accurately reproduces the behavior seen in previous experimental studies, and the peak-temperature-dependence appropriately matches experimental results with better agreement than previous theoretical descriptions[30], [31].

## 3. V[TCNE]$_x$ Material Parameters

Here we provide calculations for V[TCNE]$_x$ and YIG magnetization parameters. Accordingly, in SI units, the relations between the exchange energy $A_{ex}$, spin wave stiffness $D_{ex}$, and exchange length $\lambda_{ex}$ are given by:

$$\hbar\omega = Dk^2 = \hbar\gamma \frac{2A_{ex}}{M_S}k^2 = \hbar\gamma\mu_0 M_S \lambda_{ex} k^2 \quad (S34)$$

where $\gamma/2\pi = 28$ GHz/T, $\hbar = \frac{h}{2\pi} = 1.054 \times 10^{-34}$ J s, and $\mu_0 = 4\pi \times 10^{-7}$ N/A$^2$. Further, for a cubic ferromagnetic crystal with lattice spacing $a$, spin $S$, and number of nearest neighbors $\mathcal{Z}$, and effective exchange energy $J_{eff}$, the exchange for $k \ll \pi/a$ magnon energy is

$$\hbar\omega \approx \mathcal{Z}SJ_{eff}a^2 k^2 = D_{ex}k^2. \quad (S35)$$

Accordingly, this results in the relationship between $J_{eff}$ and $D_{ex}$ and the magnetic ordering (Curie) temperature is

$$J_{eff} = \frac{D}{a^2 \mathcal{Z}S} = \frac{3k_B T_C}{2\mathcal{Z}S(S+1)}. \quad (S36)$$

Accordingly, from previous studies[47], [49], [65] the exchange parameters and expected Curie temperatures are calculated and provided in Table S1. The small calculated $J_{eff}$ and $T_C$ are a consequence of the small exchange stiffness $D_{ex}$ and large lattice spacing $a$ of V[TCNE]$_x$. The exchange parameters far undervalue the predicted Curie temperature compared to experimental measurements [41], [62] and highlight a need for further exploration.

| $M_S$ (A/m) | $D_{ex}$ (J m²) | $A_{ex}$ (J/m) | $\lambda_{ex}$ (m²) | $a$ (nm) | $SZ$ | $J_{eff}$ (meV) | $T_C$ (K) | Ref. |
|---|---|---|---|---|---|---|---|---|
| 6093.25 | $1.34 \times 10^{-41}$ | $2.2 \times 10^{-15}$ | $0.94 \times 10^{-16}$ | 0.74 | $2 \times 6$ | 0.013 | 3.6 | 5 |
| 4774 | $5.12 \times 10^{-41}$ | $6.59 \times 10^{-15}$ | $4.7 \times 10^{-16}$ | 0.74 | $2 \times 6$ | 0.049 | 13 | 25 |

**Table S1:** Magnetic and exchange parameters for V[TCNE]$_x$. Note the $T_C$ here is calculated from the exchange and structural parameters of V[TCNE]$_x$, however these values far underestimate the magnetic ordering temperature seen in experiments.

## 3. SSE Voltage Dependence on Heater Power

The magnitude of the ISHE voltage measured in the Pt layer is directly dependent on the spin current in the V[TCNE]$_x$ layer, as in by Eq (2) of the main text. As the thermal flux $J_Q$ is increased, the thermal gradient (thus the spin current) increases accordingly. Further information about V[TCNE]$_x$ can be determined by comparing voltage measured due to the temperature gradient in the material to the heat flux and thermal conductivity. That is, consider Fourier's law of heat conduction in one dimension

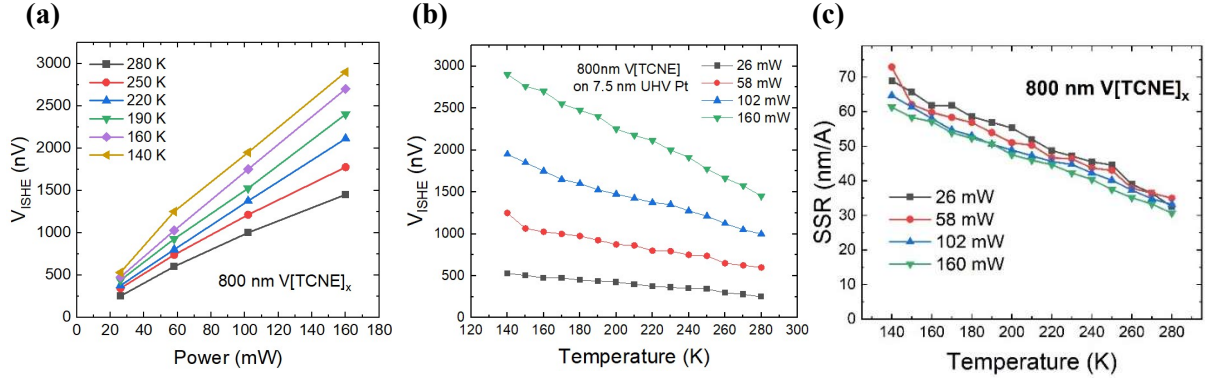

**S2:** (a) ISHE voltage from 800 nm thick V[TCNE]$_x$ film versus power for a variety of temperatures. (b) The same data in (a) with more datapoints plotted versus temperature for the applied heater powers. (c) SSR vs temperature for a variety of heater powers on the 800 nm V[TCNE]$_x$ sample, showing that the SSR is a consistent parameter with various heat fluxes.

$$\frac{dQ}{dt} = -\kappa A \frac{dT}{dz}, \qquad (S37)$$

where $dQ/dt$ is the heat flow through some area $A$, $\kappa$ is the material's thermal conductivity, and $dT/dz$ is the temperature gradient through the layer. Since $J_Q = \frac{1}{A}\frac{dQ}{dT} = \frac{P}{A}$ is the heat flux provided by the resistive heater, this shows the temperature gradient is proportional to $J_Q$ and inversely proportional to $\kappa$. Therefore, one can replace $dT/dz$ in Eq. (2) in the main text and obtain reasonable estimates for the thermal conductivity in V[TCNE]$_x$. In the theoretical calculations, a

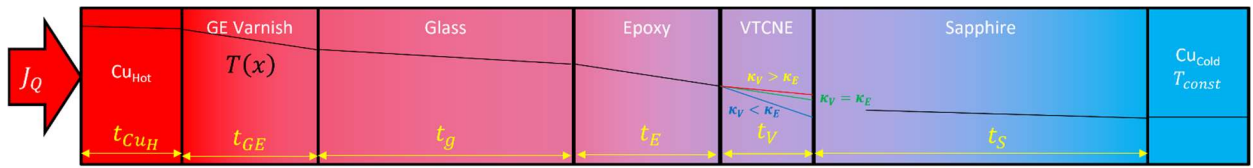

**S3:** Schematic of device layers and thermal gradients within each layer. As the heat flux is continuous through each layer, the temperature gradient in each layer is determined by $\left(\frac{1}{\kappa_i}\right)J_Q$ where $i$ denotes the layer index. That is, for high thermal conductivity materials, the thermal gradient is small (temperature nearly constant/flat, such as Cu) whereas for low thermal conductivity the gradient becomes larger. Here, the respective gradients are not drawn to scale, but are appropriate in terms of relating the magnitude of the gradients. Within the V[TCNE]$_x$ layer, the three gradients represent the potential gradients for thermal conductivity greater than (red), less than (blue), or equal to (green) the thermal conductivity of the epoxy layer. From the theory calculations, we determine via a 25 K/cm thermal gradient that an approximate value for the thermal conductivity of V[TCNE]$_x$ is 7.1 W/m·K

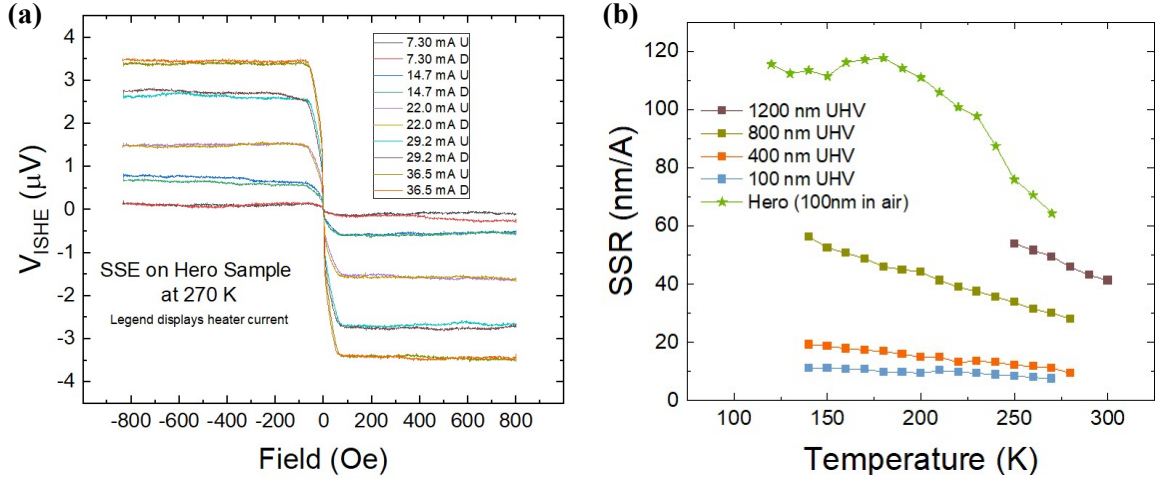

**S4:** (a) Raw data of ISHE voltage in Pt for the 100 nm thick-V[TCNE]$_x$ denoted the "Hero Sample." For this sample, Pt was deposited via e-beam evaporation and subsequently transferred in air to the V[TCNE]$_x$ growth glovebox. For unknown reasons, this sample produced an anomalously large signal that could not be reproduced in other similar thicknesses. (b) Spin Seebeck resistance of the devices shown in the main text of the manuscript compared to the Hero Sample, which shows SSR an order of magnitude larger than the 100 nm UHV-transferred sample.

value of $\frac{dT}{dz} = 25\ K/cm$ was used, consistent with thermal gradients expected in the FM layer in these devices[7], [37], therefore we find an approximate value for the thermal conductivity of V[TCNE]$_x$ of $\kappa_V = 7.1$ W/m·K which is similar to YIG's 7.4 W/m·K [59], [70].

The spin current is directly proportional to the thermal gradient in the FM layer (Eq. 2) and thus also to the input heat current provided by the heater (Eq. S34). Our data shows that the ISHE voltage is linear with applied heater power, revealing we are still in the regime where the thermal gradient (that is, energy carried by the spin current) increases linearly with heater power. As the heater power increases, the temperature of the device can change and potentially describes the deviation away from a purely linear response with heater power (see Fig. S2(a)). With decreasing temperature, the slope of the ISHE voltage versus heater power increases, corroborating the

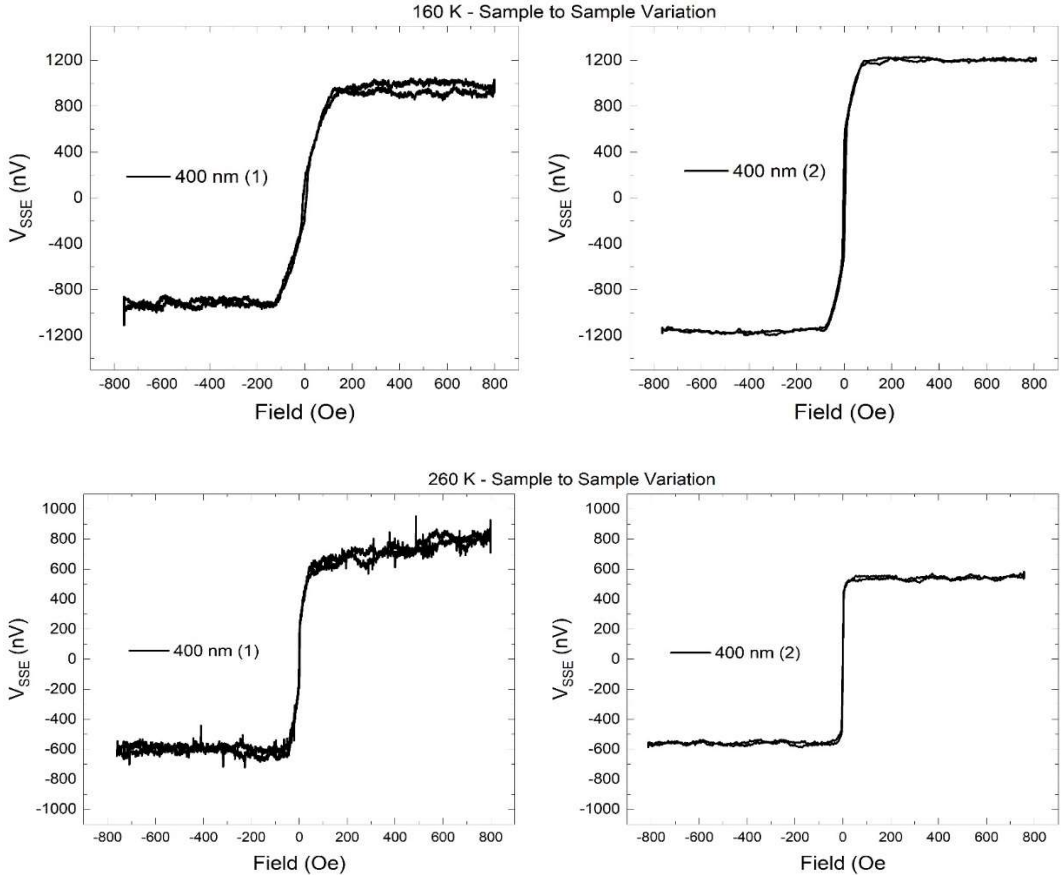

**S5:** Sample variation of ISHE voltage for nominally 400 nm V[TCNE]$_x$ samples at 160 K (top panels) and 260 K (lower panels)

prediction that the magnonic properties are dependent on temperature as discussed in the main text and above.

## 4. V[TCNE]$_x$/Pt Interfacial Quality

The interfacial quality between V[TCNE]$_x$ and Pt is paramount for observing consistent SSR signals in V[TCNE]$_x$ SSE devices. We found that the in-air transfer from the Pt deposition system to the V[TCNE]$_x$ growth glovebox resulted in significant sample-to-sample variation of SSR signals, and, as such, no other samples garnered the anomalously large SSR of the anomalous "Hero" sample shown in Figure S4. We attribute these inconsistencies to the interfacial quality between the Pt and V[TCNE]$_x$ layers. When Pt films are exposed to air, the atmospheric gases, moisture, and particulates rapidly adsorb on the Pt surface, thus producing a less-than-ideal interface upon which to deposit V[TCNE]$_x$ therefore altering the exchange coupling between V[TCNE]$_x$ and Pt. To circumvent this issue, Pt films are transferred *in-vacuo* via a UHV suitcase after MBE-Pt deposition into the V[TCNE]$_x$ growth glovebox, effectively allowing only exposure to potential surface contaminants found in the argon glovebox environment ($O_2$, $H_2O$ < 1.0 ppm, HEPA-filtered environment). Upon switching to this UHV-transfer, the sample to sample variability was significantly reduced, as seen in Supplemental Figure S5. This UHV-transfer minimizes atmospheric gas and moisture adsorption or particulate accumulation on the Pt interface and provides a reliable, near-pristine surface on which V[TCNE]$_x$ can deposit. Any further variations in sample signal sizes may be attributed to the (10-20)% uncertainty in V[TCNE]$_x$ film thickness for each growth[60].

We see the strength of the SSE decreases with temperature within the measured temperature range for UHV-Pt samples. This temperature behavior is consistent with theory and similar SSE experiments on YIG. The behavior of the anomalous Hero sample exhibits a peak around 180 K. In YIG, a similar behavior peak behavior has been seen, with the peak position moving to lower temperatures with increasing thickness as outlined in Section 2.3 above. It is possible that the same effect is present in the 100 nm UHV sample and that we are (i) not seeing a large enough signal to notice in the 100 nm UHV-Pt sample or (ii) unable to measure to lower temperatures due to sample delamination, so other modifications to the devices must be made in order to develop the full temperature profile.

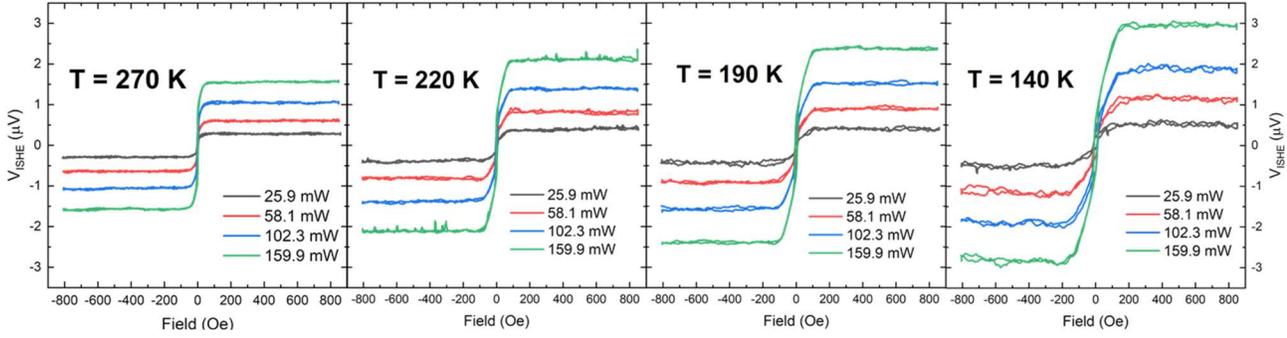

**S6:** Raw ISHE voltage sweeps for the 800 nm V[TCNE]$_x$ sample for various temperatures, showing typical behavior of SSE devices with temperature and applied heater power. Note that not only does the signal increase with decreasing temperature, but the magnetic hysteresis also becomes softer which is consistent with strain-induced changes to the anisotropy presented in the literature [Yusuf *et al.*]. This is particularly clearly shown by the variation in saturation field which increases with decreasing temperature.

## 5. Magnetic Hysteresis and Anisotropy with Temperature-Induced Strain

We notice that the magnetic hysteresis becomes softer at lower temperatures as seen above in Figs. S5 and S6, which is consistent with strain-dependent magnetic anisotropy fields in V[TCNE]$_x$ [42], [56]. The effect of this strain-dependent magnetic anisotropy on the SSE signal remains unexplored and is beyond the scope of this work. However, as the SSE is dependent on the magnetization of the film, there remains the possibility of a strain-tunable SSE in V[TCNE]$_x$, though again this remains beyond the scope of this work.